\input psfig.sty
\documentclass[12pt,preprint]{aastex}

\slugcomment{Starburst in Abell 1835.}
\shorttitle{The Starburst in Abell 1835}
\shortauthors{McNamara et al.}

\def\mathfont#1{\ifmmode{#1}\else{$#1$}\fi} %for math font     
\def\lae{\mathrel{<\kern-1.0em\lower0.9ex\hbox{$\sim$}}}  
\def\gae{\mathrel{>\kern-1.0em\lower0.9ex\hbox{$\sim$}}}  
\def\ergsec{\mathfont{ {\rm ergs\ s}^{-1}}}

\def\msun{\ifmmode{\ {\rm M}_\odot}\else{$ {\rm M}_\odot$}\fi}  
\def\msunyr{\ifmmode{\msun \ {\rm yr}^{-1}}\else{$\msun \ {\rm 
yr}^{-1}$}\fi}

\newcommand{\mdot}{ \dot{M} }

\begin{document}

\title{The Starburst in the Abell 1835 Cluster Central Galaxy: A Case Study
of Galaxy Formation Regulated by an Outburst from a Supermassive Black Hole}

\author{B. R. McNamara\altaffilmark{1}, 
        D. A. Rafferty\altaffilmark{1}}
\author{L. B\^{\i}rzan\altaffilmark{1}, 
        J. Steiner\altaffilmark{1}, 
        M. W. Wise\altaffilmark{3},
        P. E. J Nulsen\altaffilmark{2},
        C. L. Carilli\altaffilmark{4},
        R. Ryan\altaffilmark{1,5}, 
        M. Sharma\altaffilmark{1}
} 

\altaffiltext{1}{Astrophysical Institute and Department of Physics \& Astronomy, 
                 Ohio University, Clippinger Labs, Athens, OH 45701}
\altaffiltext{2}{Harvard-Smithsonian Center for Astrophysics, MS 70,
                 60 Garden Street, Cambridge, MA  02138, and University of Wollongong}
\altaffiltext{3}{Massachusetts Institute of Technology, Kavli Institute,
                 Cambridge, MA 02139--4307}
\altaffiltext{4}{National Radio Astronomy Observatory, Soccorro, NM}
\altaffiltext{5}{Department of Physics \& Astronomy, Arizona State University,
Tempe, AZ}

\begin{abstract}
We present an optical, X-ray, and radio analysis of the starburst in the 
Abell 1835 cluster's central cD galaxy. 
The dense gas surrounding the galaxy is 
radiating X-rays with a luminosity of $\sim 10^{45}$ erg s$^{-1}$ 
as its temperature ranges from $\sim 9$ keV to $\sim 2$ keV, consistent
with a cooling rate of $\sim 1000-2000 \msunyr$.  However, Chandra and 
XMM-Newton observations found less than $200 \msunyr$ of gas
cooling below $\sim 2 $ keV, a level that is consistent with 
the cD's current star formation rate of $100-180\msunyr$.  
One or more heating agents (feedback) must then be replenishing 
the remaining radiative losses.
The heat fluxes from supernova explosions and thermal conduction alone are
unable to do so.  However, a pair of X-ray cavities from an
AGN outburst has deposited 
$\simeq 1.7 \times 10^{60}$ erg into the surrounding gas over the past 40 Myr.
The corresponding jet power $\simeq 1.4\times 10^{45}~{\rm erg~s^{-1}}$
is enough to offset most of the radiative losses from the cooling gas.  The jet power exceeds the
radio synchrotron power by $\sim 4000$ times, making this one of the most radiatively
inefficient radio sources known.  The large jet power implies that the cD's
supermassive black hole accreted at a mean rate of 
$\sim 0.3 \msunyr$ over the last 40 Myr or so, 
which is a small fraction of the Eddington accretion rate
for a $\sim 10^9\msun$ black hole.  The ratio of the bulge growth 
rate through star 
formation and the black hole growth rate through accretion 
is consistent with the slope of the (Magorrian) relationship 
between bulge and central black hole mass in nearby quiescent galaxies.
The surface densities of molecular gas and star formation follow
the Schmidt-Kennicutt parameterizations, indicating
that the high pressure environment does not substantially alter
the IMF and other conditions leading to the onset of star formation. 
The cD in Abell 1835 appears in many respects to be
a textbook example of galaxy formation governed by the gravitational 
binding energy released by
accretion onto a supermassive black hole. The consistency between
net cooling, heating (feedback), and the cooling sink (star formation) 
in this system resolves the primary objection to traditional
cooling flow models.

\end{abstract}


\section{Introduction}

The most luminous galaxies in the Universe lie at the 
centers of galaxy 
clusters. Central dominant galaxies (which we refer to as cD galaxies) 
have masses of $\sim 10^{13}\msun$ and halos extending
hundreds of kiloparsecs into the surrounding cluster (Sarazin 1986). 
They are able to grow to such large sizes by swallowing stars and gas from neighboring galaxies 
(Gallagher \& Ostriker 1972, Hausman \& Ostriker 1978, Merritt 1985)
and by capturing the cooling intracluster gas
(Fabian \& Nulsen 1977, Cowie \& Binney 1977).  
The bulges of many cD galaxies are currently growing rapidly through
gas accretion and star formation proceeding at rates of $\sim 10-100 \msunyr$ 
(Johnstone, Fabian, \& Nulsen 1987, McNamara \& O'Connell 1989, 1993, 
Crawford et al. 1999, McNamara, Wise, \& Murray 2004,
Hicks \& Mushotzky 2005, Rafferty et al. 2006, this paper).  
These rates rival or exceed those found
in massive galaxies at redshifts in the range $z=2-3$ (Juneau et al. 2005), 
yet they are 
found in nearby cooling flow clusters characterized by cuspy X-ray surface
 brightness profiles and rapidly cooling gas. 
 The starbursts are fueled by $\sim 10^{8-11}~{\rm M}_{\odot}$
reservoirs of cold atomic and molecular gas 
(Jaffe, Bremer, \& Van der Werf 2001,
Jaffe, Bremer, \& Baker 2005,  Donahue et al. 2000, Falcke et al. 1998,
Heckman et al. 1989, Voit \& Donahue 1997, 
McNamara, Bregman, \& O'Connell 1990,
O'Dea, Baum, \& Gallimore 1994, Taylor 1996, Jaffe 1990, Edge 2001,
Salom{\'e} \& Combes 2003).
Bright optical emission nebulae and X-ray emission from clumps and 
filaments of gas at temperatures
of $\sim 10^7$ K and densities of $\sim 10^{-2}~{\rm cm^{-3}}$ are
a characteristic signature of these systems (McNamara et al. 2000, 
Fabian et al. 2003, McNamara, Wise, \& Murray 2004,
Crawford, Sanders, \& Fabian 2005, Jaffe, Bremer, \& Baker 2005).  
Under these conditions, the hot gas should
cool and condense onto the central galaxy at rates of several hundred to
over $1000\msunyr$ (Fabian 1994).  However, an inflow at this level
would overwhelm these galaxies with cold gas and star
formation exceeding the observed levels by factors of 10 or more. 
This implies that the cooling gas is deposited in an invisible form
of matter, or that it is condensing out of the intracluster medium
at a much lower rate.

Progress on this problem stalled for more than a decade until the 
Chandra and XMM-Newton
observatories revealed that most of the cooling gas is
not condensing out of the hot intracluster medium, but rather
it is maintained at 
X-ray temperatures by one or more heating agents.
The spectra of cooling flows fail to show the soft X-ray emission lines
emerging from gas cooling out of the X-ray band 
at the expected strength (Molendi \& Pizzolato 2001; David et al. 2001,
Fabian et al. 2001, Peterson et al. 2001, Peterson et al. 2003,
Tamura et al. 2001, B\"ohringer et al. 2002).  Instead, the gas 
seems to be cooling
down to about 1/3 of the average gas temperature at the expected rates,
but most of it fails to continue to cool and condense onto the cD galaxy
(Peterson et al. 2003).  
These observations do not, however, exclude cooling below X-ray
temperatures at levels that are comparable to the observed 
star formation rates (McNamara 2004, McNamara, Wise, \& Murray 2004, Hicks \& Mushotzky
2005, Rafferty et al. 2006).  Therefore, while the bulk of the cooling gas
remains hot, enough  may be condensing onto cD galaxies to drive
star formation and to fuel the active nucleus (AGN) at substantial rates.

Potential heating mechanisms include thermal conduction from
the hot gas surrounding the cool core (Tucker \& Rosner 1983,
Bertschinger \& Meiksin 1986, Narayan \& Medvedev 2001), 
subcluster mergers (Gomez et al 2002),
supernovae (Silk et al. 1986), and AGN outbursts 
(Tabor \& Binney 1993, Binney 2004, Soker et al. 2001,
Ciotti \& Ostriker 1997), among others.  
However, the stringent demands on these mechanisms 
have been met with varying degrees of success.  
X-ray cooling is persistent, powerful, 
and widespread.  An effective heating mechanism must be able to cope 
by producing a heat flux of 
$\sim 10^{44}-10^{45}~{\rm erg~s^{-1}}$ persisting 
over several Gyr and distributing the heat
throughout a cooling volume that is comparable to the full extent of the 
central galaxy.  Supernova explosions are generally too weak
and too localized; mergers are powerful enough but cannot
be relied upon to provide a persistent source of heat; and
conduction proceeds with great difficulty deep in the cool cores of clusters. 
Recurrent AGN outbursts have emerged as the agent best able to meet these
requirements, although thermal conduction still may play a significant 
role near the cooling radius (cf., Ruszkowski \& Begelman 2002,
Rosner \& Tucker 1989, Narayan \& Medvedev 2001,
Voigt et al. 2002, Soker, Blanton, \& Sarazin 2003, Zakamska \& Narayan 2003). 

cD galaxies are known to harbor powerful radio sources (Burns 1990).  
Outbursts from AGN associated with these sources generate 
cavities, ripples, and shock fronts in the hot gas surrounding
them, and the energy dissipated is enough
to balance radiative cooling losses in many systems 
(McNamara et al. 2000, 2001, 2005, B\^{\i}rzan et al. 2004,  
Young et al. 2002, Forman et al. 2005, Nulsen et al. 2002, 2005 a, b, 
Blanton et al. 2001, Heinz et al. 2002, Kraft et al. 2005, Rafferty et al. 
2006, and others). 
These outbursts generate $\sim 10^{55}$ erg
in giant ellipticals and groups (Finoguenov \& Jones 2001), to 
upward of $\sim 10^{61}$ erg
in rich clusters (B\^{\i}rzan et al. 2004, McNamara et al. 2005, 
Nulsen et al. 2005 a, b).  This is enough energy
to quench cooling entirely in isolated ellipticals (Finoguenov \&
Jones 2002, Best et al. 2006), and to drive 
outflows and buoyant bubbles that regulate cooling flows 
(Rafferty et al. 2006).
The most powerful outbursts are able to heat the gas 
beyond the cooling region (McNamara et al. 2005, Nulsen et al. 2005 a, b) and 
contribute to the overall entropy excess in clusters (Voit \& Donahue 2005).
The past few years have seen remarkable
growth in the number of computer simulations of pressure-confined jets
pushing through hot atmospheres. 
The simulations generally show that much of the jet
energy is thermalized and thus is able to heat the gas.  
However, the important details concerning how
and how much of the jet energy is thermalized and distributed throughout
the cooling region, and how the cavities are stabilized are not entirely
understood (e.g., Reynolds, Heinz, \& Begelman 2002, Basson \& Alexander 2002, 
Kaiser \& Binney 2003, Br\"uggen \& Kaiser
2001, 2002 Br\"uggen 2003, Soker et al. 2001, 
Brighenti \& Mathews 2002, Churazov et al. 2001, Churazov et al. 2002, 
Quilis et al. 2001, De Young 2003, Jones \& De Young 2004, 
Omma et al. 2004, Ruszkowski, Br\"uggen, \& Begelman 2004, 
Vernaleo \& Reynolds 2005, Piffaretti \& Kaastra, 2006, and many others).

These issues are deeply rooted in the more general problem of galaxy
formation.  In the standard cold dark matter
hierarchy (White and Rees 1978), small halos merge into
larger ones while the captured baryons cool and condense onto the progenitors
of mature galaxies, a process that should still be occurring
in clusters today (Cole 1991, Blanchard, Valls-Gabaud, \& Mamon 1992,
Sijacki \& Springel 2005).
This paradigm successfully describes the distribution
of matter on large scales.  However, it has difficulty dealing with
the fact that even the biggest galaxies seemed to have formed quickly.
Furthermore, CDM models that include
gravity alone over-predict the fraction of cold baryons (Balogh et al. 2001),
and thus they predict bigger galaxies and more of them 
than are observed (Voit 2005).
The jumbo cD galaxies are a case in point.  Although they have 
grown to enormous sizes, they should have absorbed more of the 
cooling baryons in clusters and grown larger still 
(Sijacki \& Springel 2005).  
Instead of condensing onto the cD, most of the baryons in clusters reside 
today in the hot gas between the galaxies.  The work-around involves
nongravitational heating by early supernova explosions and AGN. 
Supernova explosions are surely important at some level, and
they are essential for enriching the gas with metals (Metzler \& Evrard 1994, 
Borgani et al. 2002, Voit 2005).  
But they are generally too feeble and localized  
to truncate star formation in 
massive galaxies (Borgani et al. 2002).  Furthermore, in the closely related
``preheating'' problem, they have difficulty  boosting the 
entropy level of the hot gas to the 
observed levels, particularly in cooler clusters 
(Wu et al. 2000, Voit \& Donahue 2005, Donahue et al. 2005). 

A great deal of progress on these problems has been made in
recent work showing that powerful AGN outbursts in cD galaxies
can supply enough energy to reduce or quench cooling flows and 
thus regulate the growth of massive galaxies 
(e.g., B\^{\i}rzan et al. 2004 and references therein).
At the same time, lower limits on the outburst energies, which can now 
be measured reliably using X-ray
cavity and shock properties imply that the
supermassive black holes powering them are growing at typical rates of
$\sim 10^{-3}\msunyr$ (Rafferty et al. 2006).  In some cases the growth rates are approaching 
or modestly exceed $\sim 1\msunyr$ (McNamara et al. 2005, Nulsen et al. 2005 a, b) rivaling
those during the most rapid periods of black hole growth in the
early Universe.  Except in the most powerful outbursts, they
accrete at a small fraction of the Eddington rate 
(Rafferty et al. 2006), through a combination of cold disk accretion and
Bondi-Hoyle accretion of the hot gas surrounding them 
(Rafferty et al. 2006).  Bondi-Hoyle accretion is not required
as there is an adequate supply of cold 
fuel in cDs to accommodate extended periods of rapid accretion.
%The cavity properties provide relatively reliable heating rates 
%and black hole growth rates
%because the energy being liberated is overwhelmingly mechanical
%(Birzan et al. 2004, Rafferty et al. 2006, Owen et al.
%1999, Bicknell et al. XXX, De Young 199X).  The accuracy is
%limited primarily by Chandra's ability to resolve the cavities
%and not by radiation processes in the jet or the accretion disk.

Supermassive black holes may
reside at the centers of most if not all massive bulges,
and thus they appear to be an inevitable 
consequence of galaxy formation (Kormendy and Richstone 1995).
The well-known correlations between bulge luminosity, velocity dispersion, 
and  central black hole mass (Gebhardt et al. 2000, Ferrarese \& Merritt
2000) show that the growth of galaxies and supermassive
black holes are closely connected, perhaps in part through the regulation 
of inflowing gas by AGN outbursts (Begelman et al. 2005, Springel,
Di Matteo \& Hernquist 2005).
Cooling flows have emerged among the few places in the
nearby Universe where bulge and
supermassive black hole growth in massive galaxies can be examined in 
quantitative detail.  The conditions there serve as
a testbed for feedback-driven 
galaxy formation and non gravitational heating models 
(Sijacki \& Springel 2006).

%(Heinz, Reynolds, \& Begelman 1998, Reynolds, Heinz, \& Begelman
%2002, Basson \& Alexander 2002, Nulsen et al. 2002, David et al. 2001, 
%Fabian et al. 2002, Kaiser \& Binney 2003, 
%Br\"uggen et al. 2002, Br\"uggen \& Kaiser
%2002, Br\"uggen 2003, Soker et al. 2001, 
%Brighenti \& Mathews 2002, Churazov et al. 2001, Churazov
%et al. 2002, Quilis et al. 2001, De Young 2003).  An additional source

%of heat may be thermal conduction from the hot outer layers
%of clusters (Rosner \& Tucker 1989,
%Fabian, Voigt, \& Morris 2002, Narayan \& Medvedev 2001,
%Voigt et al. 2002, Soker, Blanton, \& Sarazin 2003, Zakamska \& Narayan 2003). 
%Acting in concert, a cycle of radio-induced outflow and heat 
%inflow via thermal conduction 
%(Ruszkowski \& Begelman 2002) may regulate cooling at the
%levels of cold gas and star formation observed in CDGs 
%(McNamara et al. 2000, Edge 2001,
%Edge et al. 2002). $Chandra$'s sharp images now permit the cooling 
%rates to be measured
%on the same spatial scales as star formation, providing
%a direct test of cooling and heating models.  Applying this test and exploring
%the consequences of feedback in the Abell 1068 CDG is
%the purpose of this paper.   

We examine these issues using  deep $U$ and $R$ images
of the central region of the $z=0.252$ cluster Abell 1835, 
we compare them to new and archival Chandra images, and
we examine the relationships between cooling
and star formation in the cD galaxy.
Throughout this paper, we assume ${\rm H_0}=70~{\rm km ~s^{-1}~Mpc^{-1}}$,
$\Omega_{\rm m}=0.3$, $\Omega_\Lambda =0.7$, $z=0.2523$,
a luminosity distance of 1274 Mpc, and a conversion between
angular and linear distance of 3.93 kpc per arcsec.

\section{ X-ray \& Optical Observations}

\subsection{Optical Observations}

The optical observations were obtained with the Kitt Peak
National Observatory's 4m telescope equipped with the T2KB CCD 
camera at prime focus in 1995, February.  This configuration delivered
a plate scale of $0.47$ arcsec per pixel.  
Images were exposed through the standard $U$ plus liquid
copper sulfate red-leak blocking filter, and an $R$-band
filter with an effective wavelength of $\lambda 7431$\AA\ 
that avoids contamination from strong emission lines.
Exposure times were 2100 seconds in $U$ and
1200 seconds in $R$. The target images were taken in short scan
mode, which shifts charge in the CCD during an
exposure to improve the flat field quality of the images.
The target images were individually flat-fielded using
twilight sky images; the bias level was subtracted
from each image, and they were then combined into the
science images used in our analysis.  The seeing throughout
the observations was  $\simeq 2-3$ arcsec.  The sky was transparent,
and several photometric
standard stars were observed throughout the evening. 

\subsection{Structure of the Central Galaxy}

The $R$-band image of the central $40\times 40$ arcsec ($157 \times 157$ kpc) of the cluster 
is shown in Fig. 1a.  The light is dominated by the central cluster
cD galaxy located
at RA = 14:01:02.01, Dec = +02:52:41.7 (J2000.0), 
and nearly two dozen fainter galaxies, several of which
are projected onto the cD's envelope. Fig. 1b shows the
$U$-band image after subtracting a model of the background galaxy
leaving only the blue starburst region at the center.  
The starburst is concentrated
within a nine arcsec (35 kpc) radius of the nucleus.  It is roughly circular
in projection with no obvious resolved substructure 
or tidal features.  The starburst is discussed further in Section 3.

The $U$- and $R$-band surface brightness profiles are presented in
Fig. 2.  The profiles were
constructed using elliptical annuli with fixed ellipticities of $0.22$ 
and position angles of $\simeq 331^{\circ}$, chosen by fitting
the mean values of the $R$ band isophotes beyond the starburst region.
The surrounding galaxies were removed from the images
in order to avoid contaminating the light of the cD.
The surface brightness profiles in each band were flux calibrated using
Landolt standards and transformed to the rest frame using K corrections
from Coleman, Wu, \& Weedman (1980).  The K corrections in
$U$ and $R$ are $1.176$ mag and $0.258$ mag respectively.
Fluxes were corrected for Galactic foreground extinction
using the prescriptions of Cardelli, Clayton, \& Mathis (1989),
assuming a foreground color excess of $E(B-V)=0.03$.
The profiles include statistical error bars (imperceptible in all but
the outer points),  and systematic error confidence intervals
(dashed lines) determined using the methods described
in McNamara \& O'Connell (1992).  

The halo light beyond the starburst declines according to an
$R^{1/4}$ profile.  It rises above the extrapolation of the
$R^{1/4}$ beyond 16 arcsec (63 kpc), where the
characteristic cD halo becomes visible (Schombert 1986).  
Apart from the blue core, these properties are
normal for central cluster galaxies within a redshift $z\lae 0.1$
(Porter et al. 1991).

In Fig. 3 we show the $(U-R)_{\rm K,0}$
color profile derived from the surface brightness profiles.
To establish a point of reference, the normal rest-frame colors for a
cD galaxy
generally lie within the range of $2.3-2.6$ in the inner few
tens of kpc, while the nuclear colors
tend toward the red end of this range (Peletier et al. 1990).  Fig. 3
shows that the cD's
$U-R$ color is anomalously blue in the inner 9 arcsec or so. It's central color
is approximately 1.3 magnitudes bluer than
the halo color of $(U-R)_{\rm K,0}\sim 2$ between 12 and 17 arcsec or so (47--67 kpc).
 The colors redden to a relatively normal color of
$(U-R)_{\rm K,0}\sim 2.3$ in the halo and envelope beyond 17 arcsec.
This profile is the characteristic signature of a starburst, which
we discuss in detail in Section 3.

\subsection{Chandra X-ray Observations}

The cluster was observed three times by Chandra: on December 11, 1999, 
for 19.5~ks (OBSID 495), on April 29, 2000, for 10.7~ks (OBSID 496),
and again in December, 2005 for 66.5~ks. The 66.5~ks exposure, which we
discuss in Section 4.3, is the first
part of a longer, 250 ksec ACIS-I observation being made
in pursuit of a separate project.  The 1999 and 2000 images were made with the 
back-illuminated ACIS-S3 CCD. We analyzed this data using CIAO 3.2.3 with the 
calibrations of CALDB 3.1.0. The level 1 event 
files were reprocessed to apply the latest gain and charge transfer 
inefficiency correction and filtered for bad grades. The 
light curves of the resulting level 2 event files showed
no strong flares in either observation. 
However, comparisons with XMM-Newton observations of 
Abell 1835 (Majerowicz et al. 2002, Jia et al. 2004) show that observation 495 
was affected by a mild flare (see Markevitch 2002).
This flare has the effect of significantly raising the modeled 
temperatures in the outer parts of the cluster (Schmidt, Allen, \& Fabian
2001).  We have therefore used only observation 496 
for spectral analysis. However, since the flare is likely to be 
spatially uniform, we used both observations, appropriately corrected 
for exposure, for the imaging analysis. 

The image of the combined 30~ks-equivalent exposure is shown 
in Figure 4. Outside of the core, the X-ray emission 
is fairly smooth and 
symmetrical in an elliptical distribution with an ellipticity of 
$\sim 0.12$ and position angle of $\sim 340^{\circ}$. Inside the core, 
however, the emission is more complex, with twin, off-center peaks and two 
surface brightness depressions on either side of the cD's nucleus
(see Section 4.3). 
%located at $\alpha=14:01:02.0, 
%\delta=+02:52:43.7$ (J2000). Relative to the best-fit two-dimensional beta 
%model of the inner core of the cluster emission 
%the central $3\times 3$ pixels of the north and south 
%depressions have count deficits of $20\%$ and $40\%$, respectively.
%These depressions are interpreted as AGN-blown cavities in Section
%4.3. No evidence of a non-thermal central 
%point source was found.  

To find the radial gas density and temperature distributions, 
we extracted spectra 
from observation 496 with at least 3000 counts in each
concentric annulus about the X-ray centroid with 
the ellipticity and position angle given above. The appropriate 
blank-sky background file, normalized so that the count rate of the 
source and background images match 
in the $10-12$~keV band, was used for background subtraction. 
%Response 
%files were made using the CIAO tools \emph{mkacisrmf} and \emph{mkwarf}. 
In the following spectral 
analyses, all spectra were analyzed between the energies of 0.5 and 7.0 keV
using XSPEC 11.3.2 (Arnaud 1996) and the spectra were binned 
with a minimum of 30 counts. To obtain 
temperatures and abundances, we used a model of a single-temperature 
plasma (MEKAL) plus the effects of Galactic absorption (WABS). Abell 1835 is 
located along the same line of 
sight as the Galactic North Polar Spur (Majerowicz et al. 2002), resulting 
in excess background at low energies.  However, we are interested 
only in the bright inner parts of the 
cluster where the spur's contribution to the 
background has a negligible effect on our fits. The redshift was fixed at 
$z=0.253,$ and the absorbing 
column density was fixed at the Galactic value of 
$N_H = 2.3 \times 10^{20}$~cm$^{-2}$ (Dickey \& Lockman 1990).
The temperature, abundance, and model 
normalization were allowed to vary.  To investigate 
the effects of projection and to derive electron 
densities ($n_e$), we deprojected the spectra by including the PROJCT
 model in XSPEC with the single-temperature model 
(PROJCT$\times$WABS$\times$MEKAL) and by fitting all 
spectra simultaneously. We used the deprojected temperature and 
density ($n=2n_e$) to determine the pressure ($P=nkT$), entropy 
($S=kT/n^{2/3}$), and cooling time (B\"{o}hringer \& Hensler 1989) 
of the gas in each annulus.

The profiles, shown in Figs. 5-8 are in overall agreement with the analyses 
of the same data by Markevitch (2002) and Schmidt, Allen, \& Fabian (2001),
and with the analyses of 
XMM-Newton data (e.g., Majerowicz et al. 2002, Jia et al. 2004). 
The temperature 
of the gas rises from $\sim 4$~keV in the center to $\sim 11$~keV at 
a distance of $\sim 2$ arcmin. As expected, the 
temperatures obtained from deprojection are slightly lower in the
 central regions than the projected temperatures, as the 
projected temperatures include 
emission from the hot outer parts of the cluster that is accounted 
for in deprojection. The abundance profile shows an increase 
towards the center, rising from 
approximately 1/3 of the solar abundance at a distance of $\sim 2$ arcmin 
to roughly solar abundance at the center. A spectrum extracted for the entire 
cluster within a radius of 3 
arcmin gives an average, emission-weighted temperature of 
$kT=7.8\pm 0.3$~keV and abundance of $Z=0.39\pm 0.05$~Z$_{\sun}$.

The cooling rate of the gas was estimated by adding a cooling flow model 
(MKCFLOW) to the single-temperature model 
(i.e. WABS $\times$ [MEKAL+MKCFLOW]) 
and fitting it to 
spectra extracted from the cooling region ($r_{\rm{cool}}=41$~arcsec), 
defined to be the region inside which the cooling time is 
$<7.7 \times 10^9$~yr. Fits were made to both 
a single spectrum of the entire cooling region and to spectra extracted in 
concentric, deprojected elliptical annuli. In the 
latter case, to force all cooling to be within the cooling region,
the MKCFLOW model normalization was set to zero outside the cooling
region. For each spectrum, the 
temperature of the MEKAL component was tied to the high temperature of 
the MKCFLOW component, and the MEKAL and MKCFLOW abundances were tied 
together. Parameters were fixed or 
free to vary as described above. 

We investigated several different cooling 
models. One explored the maximum cooling rate below the 
X-ray band allowed by the data.  This model was constructed 
by fixing the low temperature of the cooling flow model to 0.1 keV.
Another explored the maximum rate of cooling from roughly 
the mean gas temperature to the lowest detected temperature within 
the X-ray band by allowing the low temperature to
vary.  The first model found cooling limits as low as $\sim 30 \msunyr$, 
while the second model allowed
for cooling at rates of several thousand solar masses per year.
We also attempted to reproduce the cooling profile of Schmidt, Allen,
\& Fabian (2001) using newer calibration files, but were unable to
do so.  Because of the low exposure 
level and high particle background, it was difficult to find a stable and
robust solution to the cooling models. We arrived at the conclusion that 
Peterson's (2003) upper limit of $<200 \msunyr$ is the most reliable
measurement available, and we have adopted this value throughout the paper.

\section{The Starburst in Abell 1835}

\subsection{Star Formation Rates from Near UV Imaging}

Except where otherwise noted, our approach, methods, and rationale 
closely follow the discussion of the starburst in 
Abell 1068 (McNamara, Wise, \& Murray 2004).  
Briefly, we estimate the luminosity, 
mass, and age of the starburst first by measuring the light
emerging from the starburst population alone in the $U$ and $R$ 
bands.   This involves modeling the light profile of the older background
stellar population in each band and subtracting it from the respective image.  
The profiles generally follow an $R^{1/4}$-law beyond
the starburst, but the profiles soften considerably in the center of
the galaxy where the true shape of the background light 
is poorly known. We therefore have taken two approaches 
that effectively give lower and upper bounds to the starburst population's
flux and color.
The first involves an extrapolation of a spline fit to the
halo profile into the core of the galaxy running underneath the starburst light
(see McNamara, Wise, \& Murray 2004 and references there).  
Second, we scaled the $R$ band profile, which is minimally
affected by the burst, to fit the $U$-band light profile underneath the
burst. Both models were subtracted from the images leaving
the starburst in residual.

The ratios of the residual and model light give the fraction
of light, $f(\lambda )$,
contributed by the starburst population in each band.
These methods give $U$-band light fractions  of
25 per cent and 50 percent respectively within 9 arcsec of the nucleus.
Since the real fraction depends on the true shape of the underlying
light profile, we treat these values as lower and upper limits.
In contrast, the starburst contributes only a few percent of the
light at $R$.  The starburst population mass is then found as
$M_*=M/L(U)_*f(U)L(U)$, where
$M/L(U)_*$ is the model-dependent $U$-band mass-to-light 
ratio of the starburst
population, and $L(U)$ is its total $U$-band luminosity.

The starburst's age is estimated by comparing its  color
to two simple but representative stellar population histories
based on the Bruzual \& Charlot (2003) population models: an
instantaneous burst and continuous star formation, each of which assumes
a Salpeter initial mass function and solar abundances (the choice
of abundance has little effect on our results).  
We found the $U$-band luminosity of the starburst population alone
to be  $L(U)_*\equiv f(U)L(U)=2.6-5.9\times 10^{11}~{\rm
L_{\odot}}$, before correcting for internal extinction.
The intrinsic color of the starburst population
provides a constraint
on the population's age, history, and  mass-to-light ratio (see McNamara,
Wise \& Murray 2004).  We find a 
probable range for the starburst population's color of
$(U-R)_*\sim -0.3$ to the bluest color that is theoretically
possible $(U-R)_*\sim -1.4$.  

The blue end of the color range is broadly consistent with an instantaneous
burst that occurred less than 3 Myr ago involving a starburst
population mass of between $9\times 10^9~{\rm M}_\odot$ and  
$2\times 10^{10}~{\rm M}_\odot$.  The red end of the color range
is consistent with an aging, instantaneous burst that occurred 32 Myr
ago, or ongoing (continuous) star formation for 320 Myr.
The instantaneous and continuous starburst population masses 
are $2\times 10^{10}~{\rm M}_\odot$
and  $3\times 10^{10}~{\rm M}_\odot$, respectively.  The star formation
rate for continuous star formation over the past 320 Myr 
is $100~{\rm M}_\odot~{\rm yr^{-1}}$.  This is consitent with the spectroscopic
rate found by Crawford et al. (1999), but somewhat lower than Allen's (1995) rate.

The data are inconsistent with star formation that
has been ongoing for $\gae 1$ Gyr, as might be expected in a long-lived
cooling flow (Fabian 1994), but the measurement uncertainties
are too large to consider more complex star formation histories.  
The instantaneous burst model is always an unrealistic approximation.
However, using colors alone we cannot rule-out an intense, short-lived 
burst of star formation fueled by a rapid infusion of gas supplied,
perhaps, by a merger.
Nevertheless, the enormous amount of molecular fuel ($\sim 10^{11}\msun$)
that must be consumed by the starburst, even at the highest rates,
suggests we are  dealing with a longer-term event that is better 
described by continuous star formation extending over several hundred Myr.

The $U$-band luminosities, masses, and star formation rates above
have not been corrected for internal extinction.
Doing so reliably requires
high resolution images in two or more bands, which we do not
have.  Although no dust lanes are seen in our images, Crawford et al. (1999) 
estimated internal extinction at the level of $E(B-V)=0.38$ based on anomalous
Balmer emission-line ratios.
Correcting this effect would increase the luminosity masses
and the star formation rate above by a factor of about $1.8$,
giving a corrected star formation rate of $180~{\rm M}_\odot~{\rm yr^{-1}}$.
The reddening would also affect the population colors, leading
possibly to a somewhat younger age, which would lessen the accreted
mass somewhat. 
A star formation rate between $100-180~{\rm M}_\odot~{\rm yr^{-1}}$
is broadly consistent with our data.

\subsection{Far Ultraviolet, Infrared, and H${\alpha}$ 
based Star Formation Rates}

Other evidence for an ongoing starburst include
the detections of
$9\times 10^{10}~{\rm M}_\odot$  of molecular gas (Edge 2001),  
and a far infrared $60\mu$ flux of 
$330\pm 69$ mJy (emission at $100\mu$ is 
absent from the IRAS addscans).  The corresponding far infrared 
luminosity of  
$L_{\rm FIR}(60\mu {\rm m})= 3\times 10^{45}~{\rm erg~s^{-1}}$,
or  $\sim 10^{12}~{L_\odot}$, places Abell 1835 nearly
in the class of ultra luminous infrared galaxies.

Assuming that the the far infrared and nebular emission 
are powered by star formation, they provide independent estimates 
of the star formation rate.
Folding the infrared luminosity through Kennicutt's (1998) 
relation, we find a star formation rate of 
$138 ~{\rm M}_\odot~{\rm yr^{-1}}$, a value that lies  
midway between the dust and dust-free $U$-band estimates.

Kennicutt's relationship between H$\alpha$ luminosity and
star formation rate gives a poorer match.  Using the Crawford et al.
(1999) H$\alpha$ luminosity $5.12\times 10^{42}~{\rm erg~sec^{-1}}$,
after correcting for internal extinction and different cosmologies, 
we arrive at a star formation rate of only  $41 ~{\rm M}_\odot~{\rm yr^{-1}}$.
This is substantially lower than the U-band and infrared
continuum estimates.  However the H$\alpha$ luminosity is an
indirect and hence less reliable star formation indicator than
the ultraviolet continuum.

Using the far UV imager on the XMM-Newton
observatory, Hicks and Mushotzky (2005) found a star formation rate
of $123~{\rm M}_\odot~{\rm yr^{-1}}$, which is consistent with our rate.
As we do, they assumed a Salpeter IMF.  However, lacking color
information, they adopted a 900 Myr age for the population which is
three times the age implied by the starburst population's color.  
An age of 900 Myr implies a
color of $(U-R)\sim 0$, which is three tenths of a magnitude
redder than our measurement, but it lies at the limit of the uncertainty.
If we adopt for the moment a 900 Myr old 
population with a corresponding $U$-band mass to light ratio of $\sim 0.18$,  
we arrive at star formation rate of $50~{\rm M}_\odot~{\rm yr^{-1}}$.  
This rate is comparable to the H$\alpha$ rate, but
it lies far below the infrared and nominal $U$-band rates. 
We regard this as a tight lower limit to the star formation rate in Abell 1835.

\subsection{Abell 1835 and the Schmidt-Kennicutt Law for Star Formation}

Using a large sample of disk galaxies and infrared-selected starburst
galaxies, Kennicutt (1998) found a series of relationships
between the surface densities of both molecular gas and star formation 
and the typical orbital periods of particles within the starburst
regions.  Normal disk galaxies, the centers
of normal disks, and starburst galaxies  spanning 
a broad range of gas and stellar surface densities lie along a series
of relatively tight power-law relationships resembling the
classical Schmidt (1959) law. This level of continuity suggests 
that the gross properties of star formation, such as the
IMF, are not
strongly effected by local environmental conditions.

It would therefore be worthwhile to compare cooling flow
starbursts lying in high-pressure cluster cores
with and without strongly interacting AGN 
to see if they follow global trends.
We include in this comparison the cD in  Abell 1068
which like Abell 1835 harbors a massive starburst (McNamara,
Wise, \& Murray 2004), but unlike Abell 1835 is apparently not currently
experiencing an energetic AGN outburst.
We measured the surface densities of star formation and
molecular gas in both systems, and we calculated the orbital period
at the edge of the star formation regions following Kennicutt's (1998) 
prescription.  Our results are listed in Table 1.

The values in Table 1 follow Kennicutt's relationship between
star formation rate density versus gas surface
density $\Sigma_{\rm SFR} \propto \Sigma_{\rm gas}^{1.4}$
for starburst galaxies, normal disk galaxies, and the centers of
normal disks.  However,
they are not located among the infrared starburst galaxies,
as one might expect.  Instead, they lie among the centers of
normal disk galaxies.  This is primarily due to the large spatial extent
of the molecular gas (Edge \& Frayer 2003) and star formation regions,
which results in lower surface densities than the
infrared starbursts in Kennicutt's sample.  This is true even though 
the cooling flow star 
formation rates dwarf those of spirals.  At the same time, 
the orbital periods of stars at the edges of the cooling flow star
formation regions are between 200 and 600 Myr, which is substantially 
longer than Kennicutt's more compact starbursts.  This implies that
star formation is consuming the molecular gas before it has had
time to execute more than a few orbits, 
which is again consistent with Kennicut's parameterizations
and assumptions now routinely adopted in semi-analytical models 
of galaxy formation (eg., Kauffman 1996, Croton et al. 2006).  

The molecular gas almost certainly originated outside of the cD.
Whether it was stripped from a passing galaxy or whether it
condensed out of the cooling flow is unknown.  
In the context of the cooling flow, its mass corresponds
to  approximately $4.5\times 10^8$ years of accumulated gas
from a $200 \msunyr$ flow.  
This timescale is close to both the cooling and orbital
timescales within the starburst. The gas may have pooled following an 
interruption in a time dependent cooling flow, as
might be expected in AGN-regulated systems.  Alternatively, it may be
that the molecular gas accumulated at the center of the galaxy until 
it reached a critical density for the onset and maintenance of star formation.
If so, the apparent agreement between the cooling and star formation
rates implies that the cooling gas is feeding the reservoir of
molecular gas now at about the mean rate it has done so for the past
several hundred Myr.

It is worth noting that earlier suggestions that 
the high ambient pressure in cooling flows might
alter the Jean's unstable molecular cloud mass leading to an 
anomalous IMF (e.g. Sarazin \& O'Connell 1983) are not supported
by this analysis.  Furthermore, with growing evidence for feedback-driven 
quenching of cooling flows, it is no longer necessary to appeal
to a faint stellar repository that would justify the need
for an anomalous IMF.

\subsection{Chemical Enrichment from the Starburst}

Metal enrichment by a starburst of this size is significant
enough to enhance the metallicity of the gas in the core.  
Fig. 8 shows the
metal abundance rising from roughly 1/3 of the solar abundance at 200 kpc to
nearly solar metallicity in the starburst region.
A similar rise but with a somewhat smaller amplitude (and poorer
spatial resolution) was also seen 
in an XMM-Newton 
analysis of Abell 1835  by Majerowicz, Neuman, \& Reiprich (2002).  
Their  XMM observation
follows the metallicity profile out to about 800 kpc, where the cluster's 
average
metallicity is about 1/4 of the solar value.  The metallicity begins to
rise at a 
radius of about 160 kpc (50 arcsec), which is beyond the
edge of our profile in Fig. 8.   Following the 
procedure of Wise, McNamara, \& Murray (2004), we 
find that the starburst alone is
capable of enriching the gas in the inner 160 kpc from 1/4 of the 
solar value to
the solar value without difficulty.  In fact, the starburst is considerably
more compact than the metal-enhanced central region of the cluster.  
Were the metals produced by the
starburst confined to the star formation region, the hot
gas would become enriched to levels well above 
the solar value, which is not observed.  This problem might
be circumvented if the metal-rich gas produced by the burst
has been transported outward and mixed with the lower
metallicity gas in the halo by a merger or an outflow
driven by the AGN.  Alternatively, the metals produced in
the starburst may be primarily locked in the cold gas and
are unable to enhance the metallicity of the hot gas.  
Finally, the observed metallicity gradient 
could have been imprinted by stellar evolution of the
cD's older population (De Grandi et al. 2004).  
This issue will be explored further in a future paper.

\subsection{Comparison Between the Cooling Rate and Star Formation Rate}

The starburst in Abell 1835 sits in a region where
the cooling time has fallen below $6\times 10^8~{\rm yr}$,
which is close to the age of the starburst.  The 
temperature of the gas has also reached a minimum
of 3.5 keV there, which is similar to other cooling flow systems, 
such as Hydra A (McNamara et al. 2000) and Abell 1068 
(McNamara, Wise, \& Murray 2004).  
These conditions are qualitatively consistent with expectations
for fueling by the cooling flow.  However, quantitative consistency
requires that there be mass continuity between the rate of cooling
and its sink.  The vast gulf between radiative cooling rates
and star formation cast a pall on the pure (no feedback)
cooling flow model since it was conceived nearly 30 years ago.  

An observation
made with XMM-Newton's reflection grating spectrometer (Peterson et al. 2003)
shows that the gas in Abell 1835 is cooling at a rate of between 
$1000-2000~{\rm M}_\odot~{\rm yr^{-1}}$ from the mean gas temperature 
of nearly 9 keV to about 2 keV where cooling slows dramatically 
(Peterson et al. 2003).  Below 2 keV, cooling proceeds at a much
reduced rate of 
$\lae 200~{\rm M}_\odot~{\rm yr^{-1}}$.  Whether 
any cooling out of the X-ray band occurs
is still to be demonstrated.  However, cooling at this 
reduced rate is now consistent with the star formation rate 
in the cD galaxy.  Provided there is an active heating mechanism
to offset the remaining cooling luminosity,
which we justify below, the (reduced) cooling model is 
now consistent with a sink in star formation and the central
black hole.  This situation is evidently true in an increasing fraction
of cooling flow clusters (Rafferty et al. 2006).

\section{Feedback \& Regulated Cooling in the Cluster's Core}

The data discussed in the previous section are consistent
with an active but relatively moderate cooling flow.
Nevertheless, the gas throughout the central 150 kpc of the cluster
has a  cooling time that is shorter than the cluster's age (Fig. 7),
yet most of this gas is not condensing out. 
The radiation losses from this gas,

%by a robust heating source 
%s conduction or the AGN (Soker et al. 2000, 
%Fabian et al. 2001, B\"ohringer et al. 2001, Binney 2004).  

\begin{equation}
L_{\rm cool}\simeq 1.2 \times 10^{45}~\left({\dot M \over 1000 \msunyr}\right)~{\rm erg~sec}^{-1},
\end{equation}

\noindent
must then be replenished by  heating 
(Fabian et al. 2001, B\"ohringer et al. 2001).
We now consider whether thermal conduction,
the AGN, and supernova explosions are able to 
compensate the radiative losses.  

\subsection{Feedback from the Starburst}

Here we adopt an optimistic star formation rate of 
$180 ~{\rm M}_\odot~{\rm yr^{-1}}$, and we follow closely the
discussions in McNamara, Wise, \& Murray (2004), 
and Wise, McNamara, \& Murray (2004). We
assume a blast energy per type 2 supernova of $10^{51}~{\rm erg}$,
and a type 2 supernova production rate of one per $127~{\rm M}_\odot$ of
star formation (Hernquist \& Springel 2003).  These values then yield 
an average energy
injection rate over the life of the starburst 
of $4.5\times 10^{43}~{\rm erg~s^{-1}}$.  
This is at most a few percent of the power required to 
quench the cooling flow, even with 
efficient coupling between the supernova blast energy and the
hot gas.  This figure can be boosted by adopting an extreme supernova 
yield per mass, or an IMF richer in massive stars than 
the Salpeter function.  However, the requirements are
still extreme in view of earlier arguments weighing against an IMF 
that is dramatically
different from Salpeter's.  Supernovae may be an important
source of heat in the region surrounding the starburst
and AGN,  but 
they cannot have a substantial effect on the overall cooling flow.
The same conclusion was reached for the Abell 1068 cluster cD
(McNamara, Wise, \& Murray 2004).  Since both are among
the most massive 
starbursts known in cooling flows, this conclusion
probably holds in most systems.

\subsection{Heating by Thermal Conduction}

The conditions in which inward-flowing heat from 
the hot layers of gas surrounding the cooling core
are able to replenish radiation
losses have been studied extensively in recent years
(Fabian, Voigt, \& Morris 2002, Zakamska \& Narayan 2003, Voigt \& 
Fabian 2004).  Abell 1835 has been examined in this context
but with contradictory results.
Zakamska \& Narayan (2003) were apparently able to construct theoretical 
gas temperature, density, and cooling profiles that matched those
of Abell 1835 using an inward heat flux proceeding
at 40 per cent of the Spitzer rate.
On the other hand  Voigt \& Fabian (2004) found that heat conduction proceeding
at a modest fraction of the Spitzer rate could quench cooling only in 
the outer reaches of the cooling region,  but not near the central
starburst where the gas temperature is rapidly decreasing. 
In order to balance radiative losses, they found that the
conductivity must exceed the Spitzer value within the radius at which
the gas falls below 7 keV.  The conductivity reaches 1/3 of
the Spitzer value where the gas temperature is approximately
10 keV.  The corresponding
radii are 20 arcsec (79 kpc) and 40 arcsec (157 kpc), respectively. 
It seems reasonable then to expect thermal conduction to balance
radiation losses in the outer parts of the cooling region, but to be
unable to do so in the central region near the starburst.
Although similar conclusions were reached for the Abell 1068 cluster 
(Wise, McNamara, \& Murray 2004),  the importance of
conduction without knowledge of the
conductivity of the gas is difficult to evaluate with confidence.

\subsection{AGN Feedback: X-ray Cavities}

The structure in the inner 10 arcsec of the X-ray image 
shown in Fig. 9 was first reported by 
Schmidt, Allen, \& Fabian (2001), who attributed it, we now
believe incorrectly, to a recent merger.  Two 
surface brightness depressions with count deficits of
approximately 20\% to 40\% compared to the surrounding
emission are seen in the ACIS-S images  
6 arcsec (23 kpc) to the north-west and 5 arcsec (17 kpc) to
the south-east of the cD's nucleus.  The cavities were
confirmed by the  66.5 ksec ACIS-I image shown at left in Fig. 9.  
The nucleus lies in the trough between the two bright central knots of
emission.  The trough might be caused by photoelectric absorption of
X-rays by the molecular gas clouds.
The cavities have bright rims and otherwise
resemble the AGN-induced X-ray cavities now seen in more than
two dozen clusters (Rafferty et al. 2006, B\^{\i}rzan et al. 2004,
Dunn \& Fabian 2005).  Their physical characteristics are 
given in Table 2, including their distances from the nucleus, 
the sizes of their major and minor axes, the surrounding gas pressures 
and deprojected temperatures, the $1pV$ energies of
the cavities, and the approximate buoyancy ages 
(see B\^{\i}rzan et al. 2004).

We were initially concerned that the cavities themselves might be caused by 
photoelectric absorption
from the molecular gas in the cD.  Edge \& Frayer (2003) found a 
column density of $4\times 10^{22}~{\rm cm}^{-2}$ in the 
inner 10 kpc region of the cD, which is centrally
concentrated and does not correspond to the
two off-axis surface brightness depressions. 
We found a small excess column density of 
$\sim 3\times 10^{21}~{\rm cm}^{-2}$ from the X-ray spectrum of
the inner 10 arcsec or so, as do Schmidt, Allen, \& Fabian (2001), which
could be the diluted column of molecular gas.  However, because the
cavities are seen in both hard and soft images above and below
2 keV, they cannot be due to photoelectric absorption,
which dominates below 2 keV.

The sizes of the cavities are more or less
typical of those found in massive  clusters
(e.g., B\^{\i}rzan et al. 2004, Dunn \& Fabian 2004). However, 
the central pressure of 
$1.1\times 10^{-9}~{\rm erg~cm}^{-3}$ is 
more than an order of magnitude larger
than is typically found at the base of a cooling flow. 
The work required to inflate the cavities against the surrounding
pressure is $pV=4.3~ (-1.5,+4.6)\times 10^{59}~{\rm erg}$.  
This corresponds to a mean mechanical power of 
$L=3.5~ (-1, +3.0)\times 10^{44}~{\rm erg~s^{-1}}$, assuming a rise time of about
40 Myr.  The total enthalpy
is roughly  $2.5$ to 4 times larger, depending on the equation of state
of the gas filling the cavity (B\^{\i}rzan et al. 2004).  This implies a total jet power of
$\sim 1.4\times 10^{45}~{\rm erg~s^{-1}}$, which is comparable to the cooling
luminosity of a $1000-2000\msunyr$ cooling flow. So long as the coupling
between the AGN and the gas is
reasonably efficient, the AGN power is high enough to offset radiation losses
in this system.

%(eg., Churazov et al. XXXX, Omma \& Binney XXXX,
%Ruszkowski \& Begelman, Soker et al. 2000,
%-Fabian et al. 2001, B\"ohringer et al. 2001, Binney 2004),
%and supernova explosions
%in the starburst itself are able to replenish the radiative losses.

The cD harbors a compact radio source shown in Fig. 9.
The image obtained from the Very Large Array archive
was taken in the A configuration at a frequency of 1.4 GHz.
The resolution of the image is about one arcsec. 
The spectral index 
$\alpha =0.65$ ($f_\nu \propto \nu^{-\alpha}$) implies 
a total radio luminosity of
$3.55\pm 0.09 \times 10^{41}~{\rm erg~s^{-1}}$ between
10 MHz and 10 GHz. (A more complete discussion of the radio properties
will be given in a forthcoming paper.)
The radio source at this frequency shows no obvious connection to the cavity
system; however, the 330 MHz map (also discussed in a forthcoming paper) 
covers the entire extent of the cavities, but again
shows no detailed correlation with the holes.  The 
low frequency source is probably the remnant synchrotron
emission from the outburst which occurred about 40 Myr
ago.  

The nucleus of Abell 1835 is a striking example of how poorly
radio synchrotron emission traces true jet power. 
The average jet power required to inflate the cavities,
$\sim 1.4 \times 10^{45}~{\rm erg~s^{-1}}$,
dwarfs the total radio synchrotron power, exceeding it by 
a factor of 4000.  
The corresponding synchrotron radiative efficiency is then 
only about $0.02\%$, 
which is vastly smaller than $\sim 1\%$ found for M87 (Owen et al. 2000)
and other sources (De Young 1993, Bicknell, Dopita, \& O'Dea 1997).  
 
\section{Simultaneous Growth of the Bulge and the Supermassive Black Hole}

The magnitude of the AGN outburst implies that the central
black hole has accreted $\simeq 4pV/\epsilon c^2 = 1.1 \times 10^7~{\rm M_\odot}$ ($\epsilon = 0.1$) in the past 40 Myr or so, corresponding to an
average accretion rate of 
$\sim 0.3 ~{\rm M}_\odot~{\rm yr^{-1}}$ (Rafferty et al. 2006).  
Adopting the star formation rate as an estimate of the 
current bulge growth rate
(see Rafferty et al. 2006 for a more detailed discussion), we find
that the bulge has added between 300 and 600 units of mass
for every one unit that has fallen into the black hole.
This relative growth rate is intriguingly
close to the slope of the Magorrian relation
between bulge and black
hole mass in quiescent galaxies (H\"aring \& Rix 2004).  
The convergence of several factors, including the fact that 
star formation in the cD is proceeding at a rate that rivals or
exceeds those observed during the peak years of galaxy
formation (Juneau et al. 2005),
suggests that the physical conditions driving this growth could be
analogous to those that held in the early universe
when the Magorrian relation was presumably imprinted on galaxies.
The cooling flow systems probably differ, however, in 
that their accretion is substantially sub-Eddington.  The conditions
in Abell 1835 should be placed in context with similar systems,
which clearly show a trend between star formation and
black hole growth, but there is a 
great deal of scatter (Rafferty et al. 2006).  
In some systems the black hole has grown by a substantial fraction 
of its mass in a single outburst but with little commensurate
bulge growth over the same time period 
(McNamara et al. 2005, Nulsen et al. 2005 a).  In other systems,
such as Abell 1068 (McNamara, Wise, \& Murray 2004), the bulge is 
growing much faster than the black hole. In general,
cooling flow cDs and their black holes 
have evidently not grown in lock-step over the past several
Gyr (Rafferty et al. 2006).

%This is not necessarily a problem. The mass of 
%the central black hole can be crudely estimated by folding the K-band 
%luminosity of the cD through the local scaling between bulge luminosity 
%and black hole mass gives a mass of $5.5\pm 3 \times 10^9~{\rm M}_\odot$. 
%Therefore, the central black hole has probably grown by only about 0.3\% 
%of its probable size during the current AGN outburst. This is true
%in general in these systems.  

\section{Summary \& Discussion}

The cD galaxy in Abell 1835 is in the midst of
a starburst proceeding at a rate of $100-180 \msunyr$
that began approximately $320$ Myr ago. 
The star formation rate is consistent with
the maximum rate that gas can be condensing out of the cooling flow.
Cooling and accretion at this level can account for
only $10\%$ to $20\%$ of the total radiative losses, implying
that the bulk the gas is being heated and maintained at X-ray temperatures.
Supernovae in the starburst are energetically incapable of producing 
enough heat to do so, and 
thermal conduction is ineffective in the inner regions of
the cooling flow.

We discovered a pair of cavities in the hot
gas produced by a powerful AGN outburst that occurred roughly 40 Myr ago.
The outburst was energetic enough to offset the remaining radiative
losses.  There is no longer a discrepancy between the radiative cooling
rate and the sink of the cooling gas, provided the jet power heats the
gas efficiently.   The jet power required to produce the cavities
exceeds the radio synchrotron power by $\simeq 4000$ times, indicating
a radiatively inefficient yet powerful radio source.

The jet power $1.4\times 10^{45}~\ergsec$
corresponds to the Eddington luminosity of
a $\sim 10^7 \msun$ black hole.  However, the K-band luminosity of
the host cD galaxy implies a much larger black hole mass of
approximately $5\times 10^9\msun$ (Rafferty et al.
2006), implying that accretion is proceeding at a small fraction
$\sim 3\times 10^{-3}$ of the Eddington rate.  This
exceeds the Bondi rate by nearly 500 times, assuming
the measured central gas density holds near the unresolved Bondi radius
(Rafferty et al. 2006). Minding the uncertain assumptions
about the black hole mass and surrounding gas density, Bondi accretion
could contribute at some level, but is unlikely to be primarily
responsible for feeding the outburst.  The large
pool of centrally condensed molecular gas is consistent with cold accretion.
If the molecular gas is fed by gas condensing out of the hot phase,
it would provide a natural supply of fuel necessary to maintain an AGN feedback
loop.  We cannot, however, exclude the possibility that the gas 
arrived through a merger.

The rate of black hole growth implied by the jet power
and bulge growth through star formation are consistent with the slope of
the (Magorrian) relationship between bulge mass and black hole mass
for quiescent bulges.  This surprising result suggests that
feedback processes like those operating in this system could be
driving the relationship between bulge mass and
supermassive black hole mass in normal bulges 
(Magorrian et al. 1998, Kormendy \& Richstone
1995, Springel DiMatteo \& Hernquist 2005).
In a large sample of cooling flows, Rafferty et al. (2006) have
found a trend between bulge growth rate
through star formation and black hole growth
rate over the past $\sim {\rm Gyr}$.  However, the large scatter in
the relative rates implies that bulge and black hole
growth do not always proceed in lock-step.
This result also supports the growing consensus that AGN 
feedback could explain
the exponential decline in luminous galaxies relative to the
predicted shape of the dark matter halo mass function by suppressing
cooling (Benson et al. 2003, Croton et al. 2006, Best et al. 2006).
Consequently, the mode of accretion in this and other cooling flow
systems should hold considerable interest to more general models
of galaxy formation (e.g., Croton et al. 2006, 
Soker 2006, Pizzolato \& Soker 2005, Sijacki \& Springel 2006).

Abell 1835's nuclear outburst is as powerful
as a quasar's and its starburst is proceeding at a rate
that rivals those in burgeoning galaxies beyond $z=2$.  
However, there are noteworthy differences in the way the gravitational
binding energy of accretion is channeled away from the black hole.  
While quasars radiate away most of
their accretion energy,  the accretion energy emerging from
Abell 1835 and other cooling flows is almost entirely mechanical. 
Abell 1835's jet power $\sim 1.4\times 10^{45}~\ergsec$ exceeds 
postulated protogalactic wind luminosities (Silk \& Rees 1998)
by an order of magnitude, as it must to all but stop the cooling flow.
However, strong nonthermal nuclear emission and broad emission
lines are absent, and the radiative efficiency of the radio source, 
defined as the
ratio of radio synchrotron power to jet power, is very low.
This may be a characteristic of the late stages of galaxy formation
when accretion onto the black hole falls
below the Eddington rate which probably held during the early stages 
of galaxy formation
(Silk \& Rees 1998, Blandford 1999, Begelman \& Nath 2005, Churazov et al.
2005).
Even so, long-lived accretion at the present rate would continue to
drive star formation and black hole growth 
such that the relationship between bulge and
black hole mass found in quiescent ellipticals would be imprinted or
maintained.

Following a checkered and often contentious history, the cooling flow
problem is now coming to resolution.
We should emphasize, however, that the model must still pass
an essential test.  Gas cooling out of the intracluster medium should emit
detectable X-ray cooling lines, notably the
Fe XVII line at 15\AA~(0.826 keV). In this case, the flux scales as
\begin{equation}
f(Fe XVII) = 5.456 \times 10^{-15} \left( \frac{\mdot}{100 \msunyr} \right)
\left( \frac{D}{500 ~{\rm Mpc}} \right)^{-2} ~{\rm ergs} ~{\rm s}^{-1} ~{\rm cm}^{-2}
\end{equation}
where $D$ is the distance and $\mdot$ is the cooling rate.
Peterson's et al (2003) limits successfully ruled-out wholesale cooling
but they generally lie well above the levels of accretion implied
by the observed star formation rates (Rafferty et al. 2006).  
A cooling rate $\mdot$ approximately equal to the observed star formation rates
is currently detectable using deep XMM-Newton observations
and will be in the future using shorter observations with
the Constellation-X observatory.
For the half dozen or so objects whose star formation rates 
are comparable to the available cooling upper limits 
(including Abell 1835 and Abell 1068), a $200-400$ ksec XMM observation
using the reflection grating spectrometer would be 
sufficient to restrictively test the cooling flow/feedback model
of galaxy formation.

\acknowledgments

This research was funded by NASA Long Term Space Astrophysics
Grant NAG4-11025.

\clearpage

%% the file, uncomment the \plotone, \plottwo, and
%% If you need a layout that cannot be achieved with \plotone or
%% \plottwo, you can invoke the graphicx package directly with the
%% \includegraphics command or use \plotfiddle. For more information,
%% please see the tutorial on "Using Electronic Art with AASTeX" in the
%% documentation section at the AASTeX Web site,
%% http://www.journals.uchicago.edu/AAS/AASTeX.
%%
%% The examples below also include sample markup for submission of
%% supplemental electronic materials. As always, be sure to check
%% the instructions to authors for the journal you are submitting to
%% for specific submissions guidelines as they vary from

%% This example uses \plotone to include an EPS file scaled to
%% 80% of its natural size with \epsscale. Its caption
%% has been written to indicate that additional figure parts will be
%% available in the electronic journal.

\clearpage

%% Here we use \plottwo to present two versions of the same figure,
%% one in black and white for print the other in RGB color
%% for online presentation. Note that the caption indicates
%% that a color version of the figure will be avaquench wholesalequench wholesaleilable online.
%%

\begin{figure}
\epsscale{1.5}
\plottwo{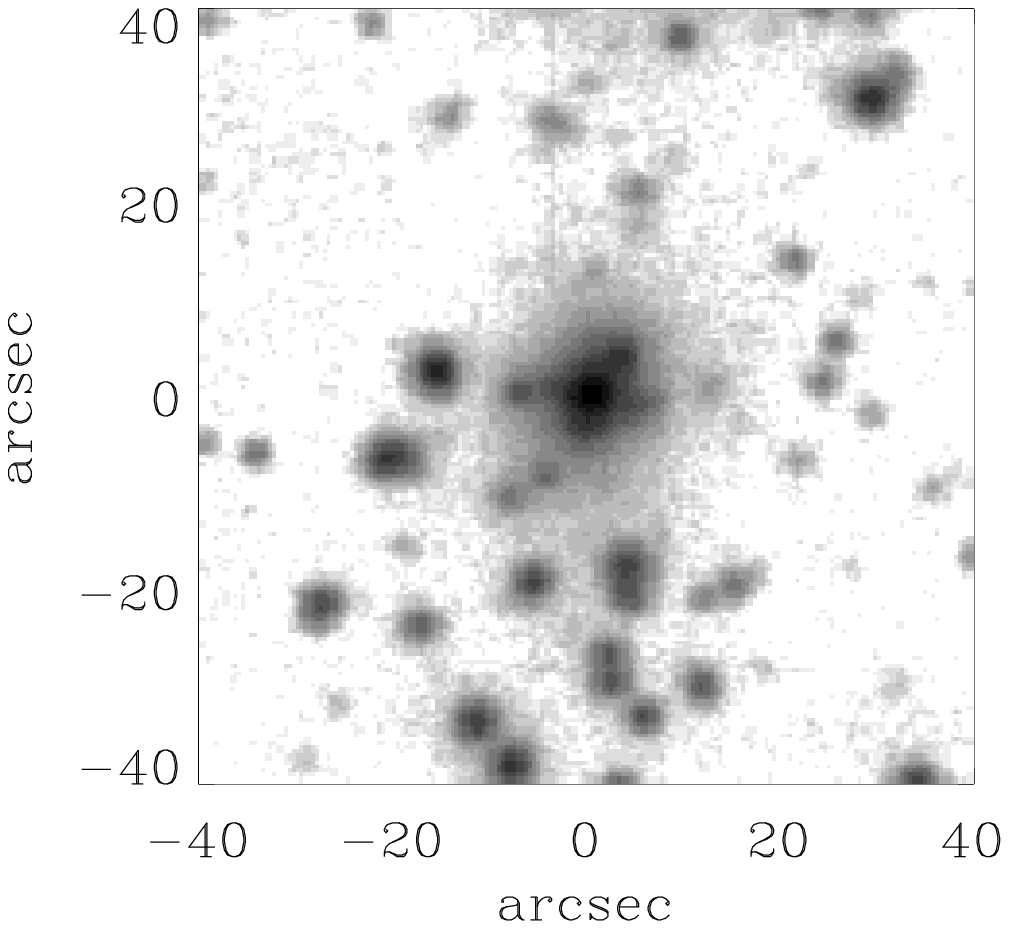}{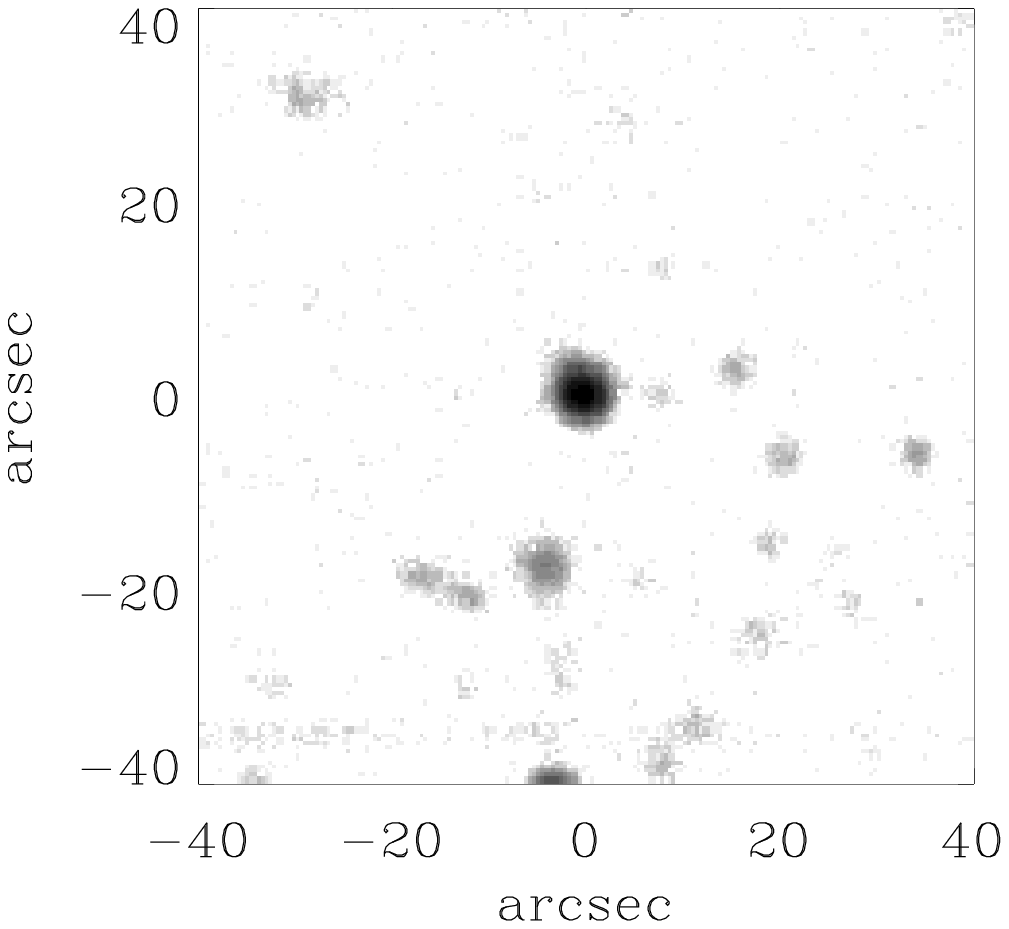}
\caption{Fig. 1- {\bf a} R-band image of the $40 \times 40$ arcsec 
($157\times 157$ kpc)
region of the cluster centered on the cD galaxy. 
{\bf b} U-band image of the same region, but
with the background population of the cD removed showing 
the central starburst.}
\end{figure}

\clearpage

\begin{figure}
\epsscale{.80}
\plotone{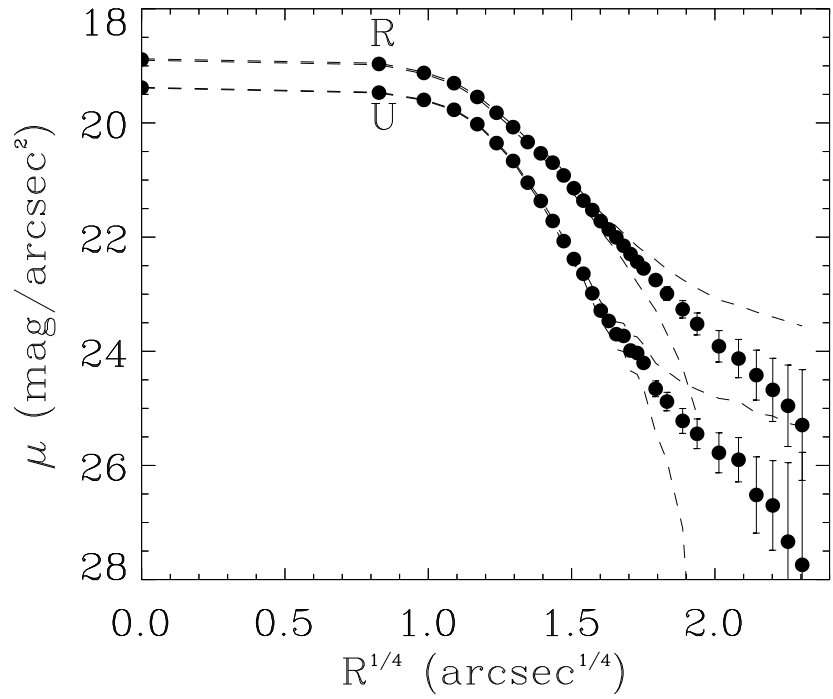}
\caption{\label{fig2} U-band and R-band surface brightness profiles of the cD galaxy. The dashed lines represent the systematic uncertainty associated with sky background subtraction}
\end{figure}

\clearpage

\begin{figure}
\epsscale{.80}
\plotone{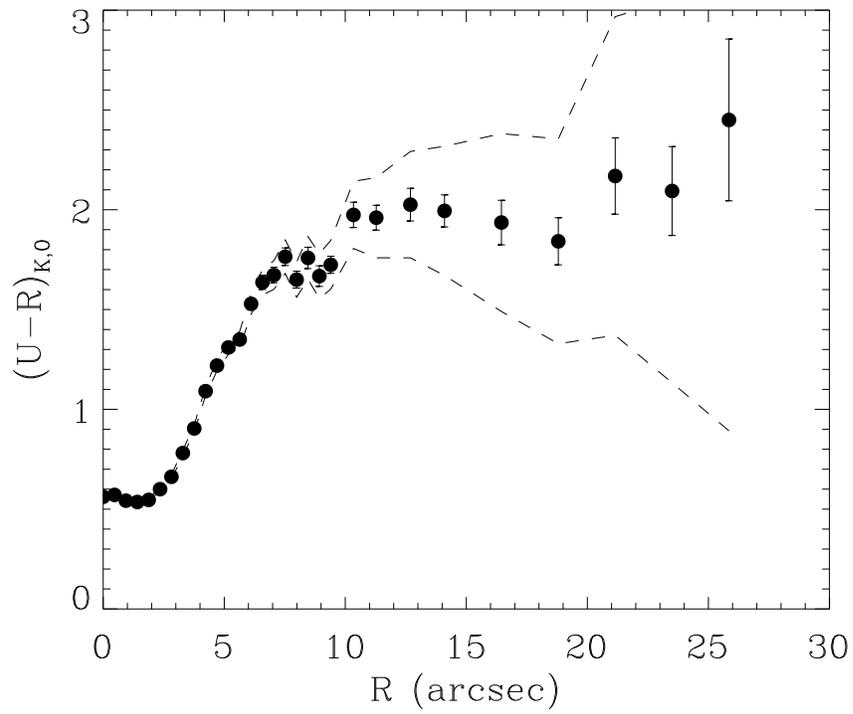}
\caption{ \label{fig3} $U-R$ color profile of the cD showing the central blue colors
associated with the starburst. Normal colors are $\sim 2.3-2.6$. The dashed lines represent the systematic uncertainty associated with sky background subtraction}
\end{figure}

\clearpage 

\begin{figure}
\plotone{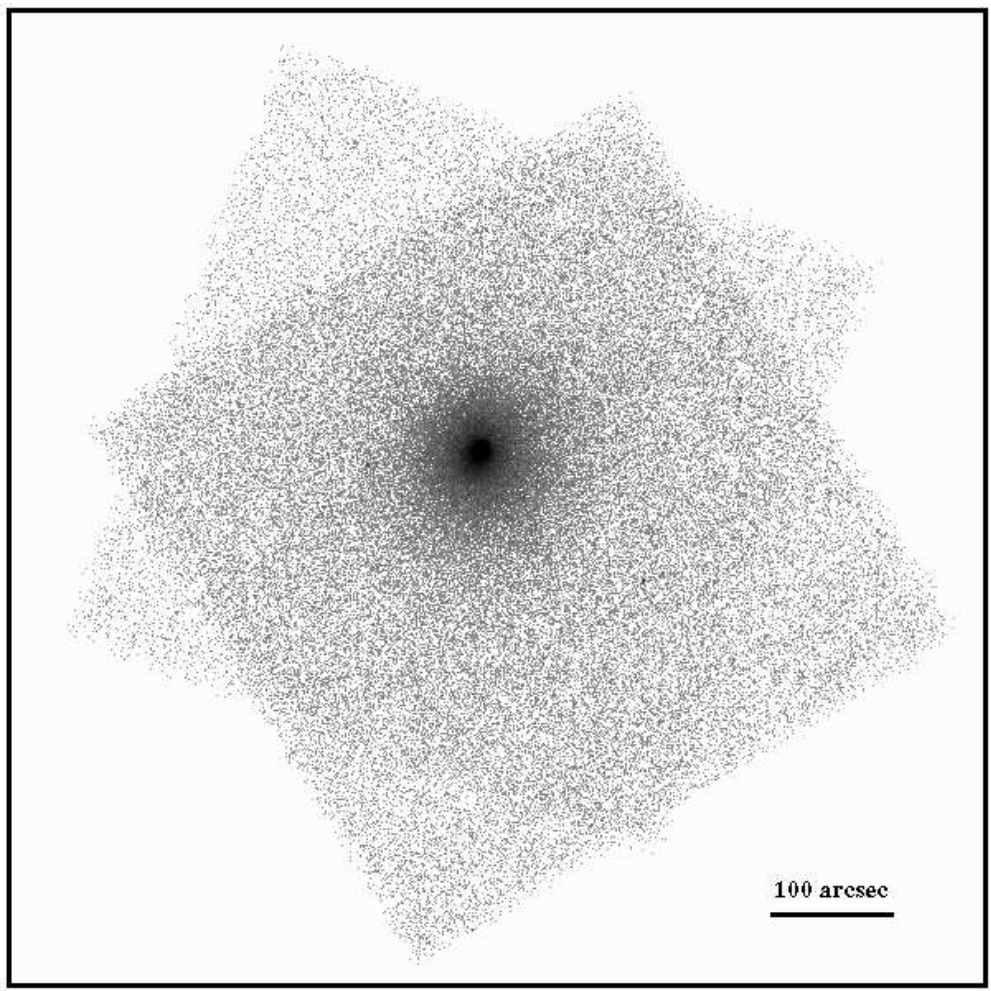}
\caption{X-ray image of the cluster made with the combined 30-ks exposure.  
\label{fig-xray-image}}
\end{figure}

\clearpage

\begin{figure}
\plottwo{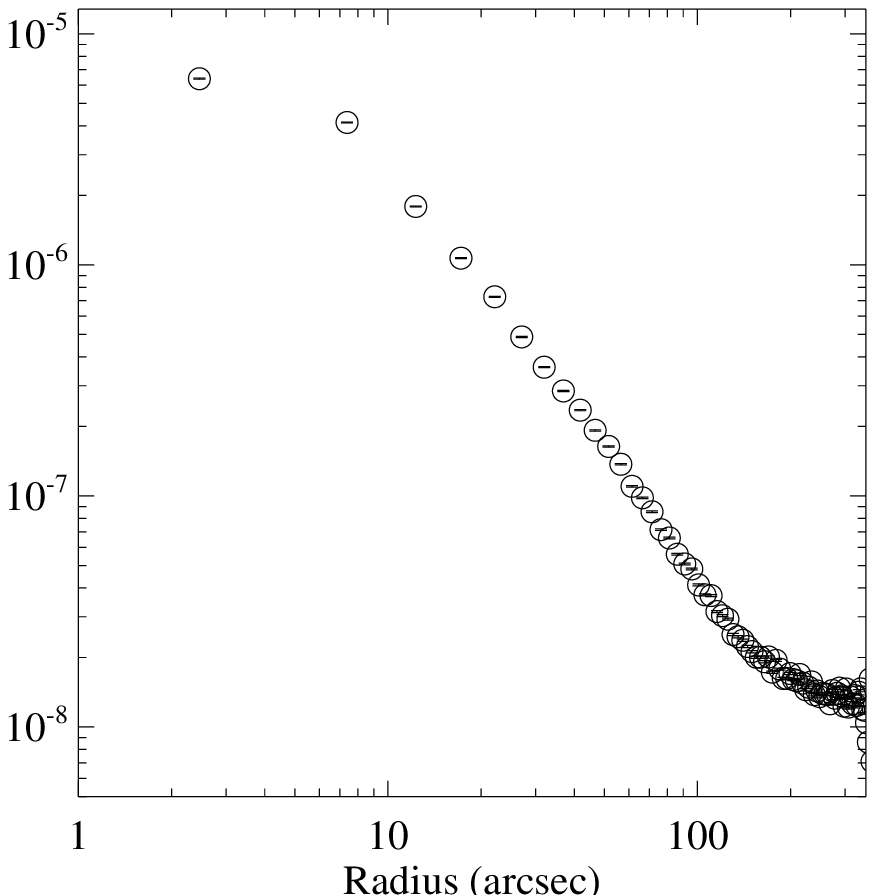}{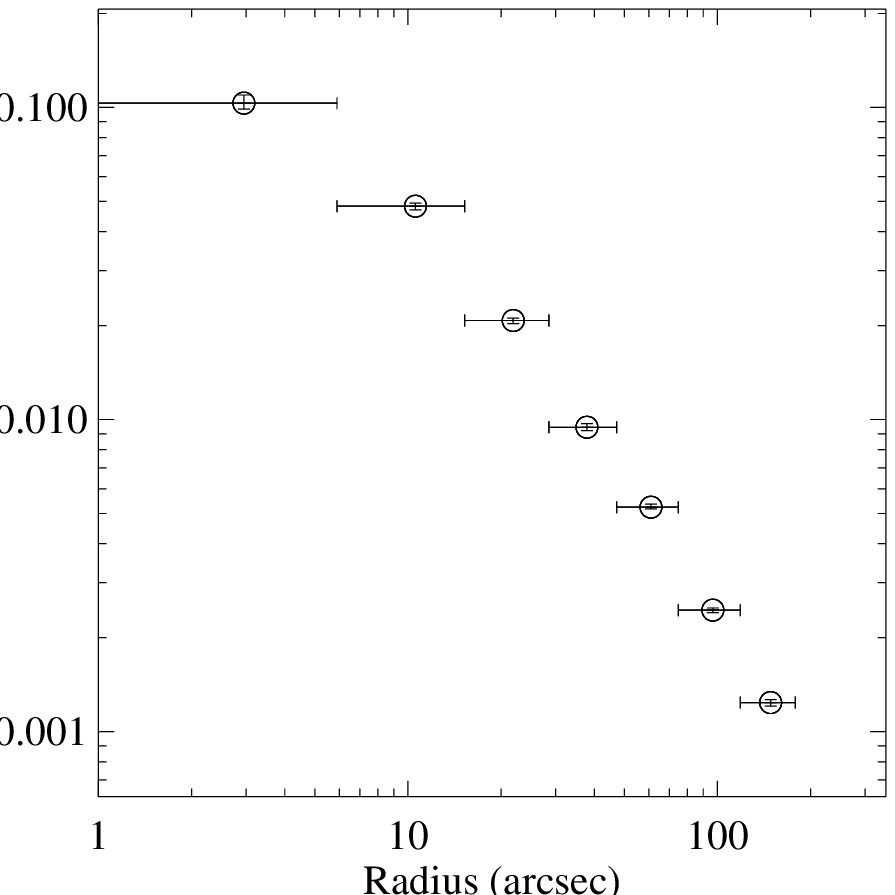}
\caption{\emph{Left:} X-ray surface brightness profile.  \emph{Right:} Density profile of the hot gas.. 
\label{fig-xray-image}}
\end{figure}

\clearpage

\begin{figure}
\plottwo{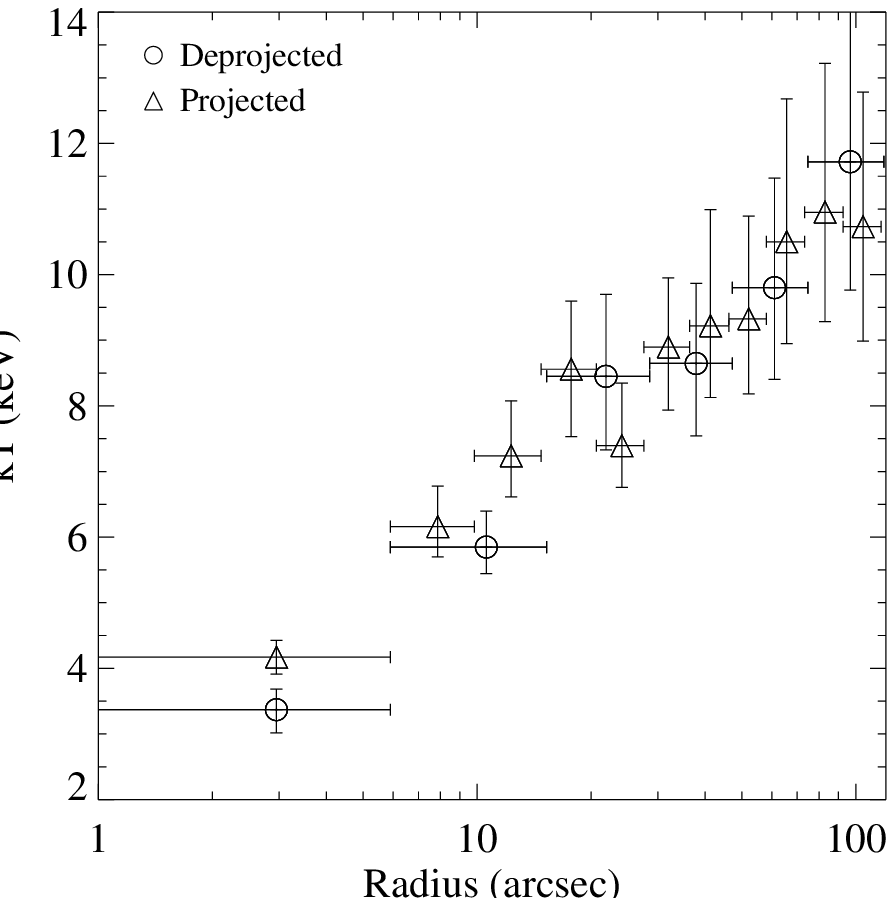}{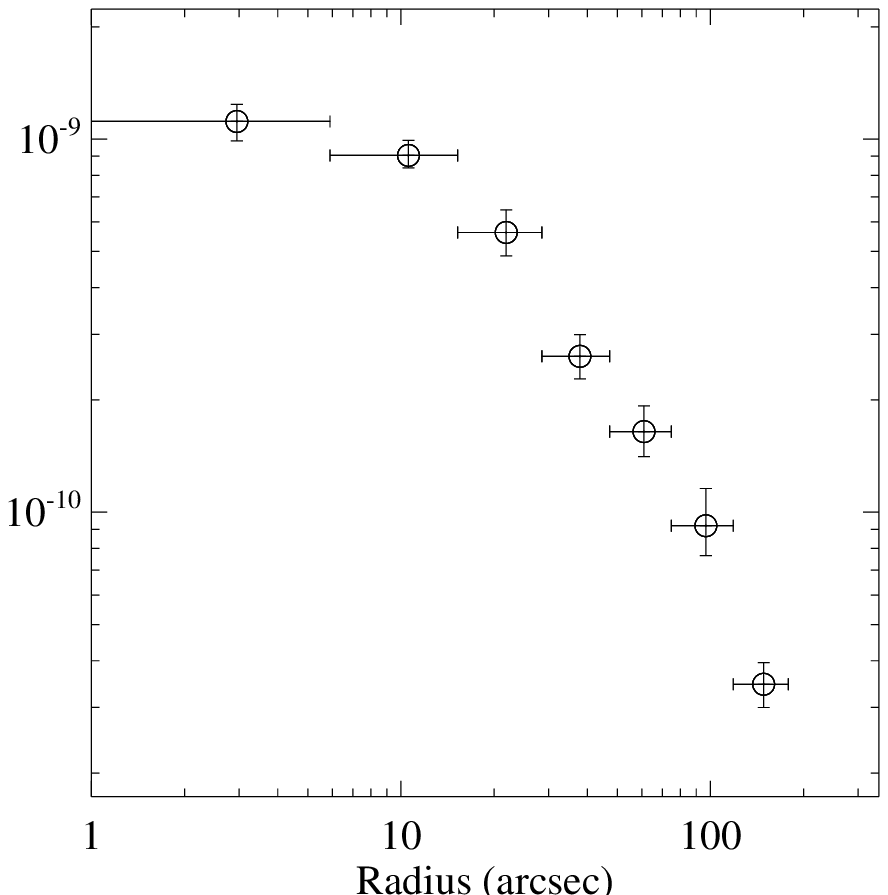}
\caption{\emph{Left:} Projected (triangles) and deprojected (circles) 
X-ray temperature profile of the hot gas.  \emph{Right:} Central pressure profile of the hot gas.
\label{fig-xray-image}}
\end{figure}

\clearpage

\begin{figure}
\plottwo{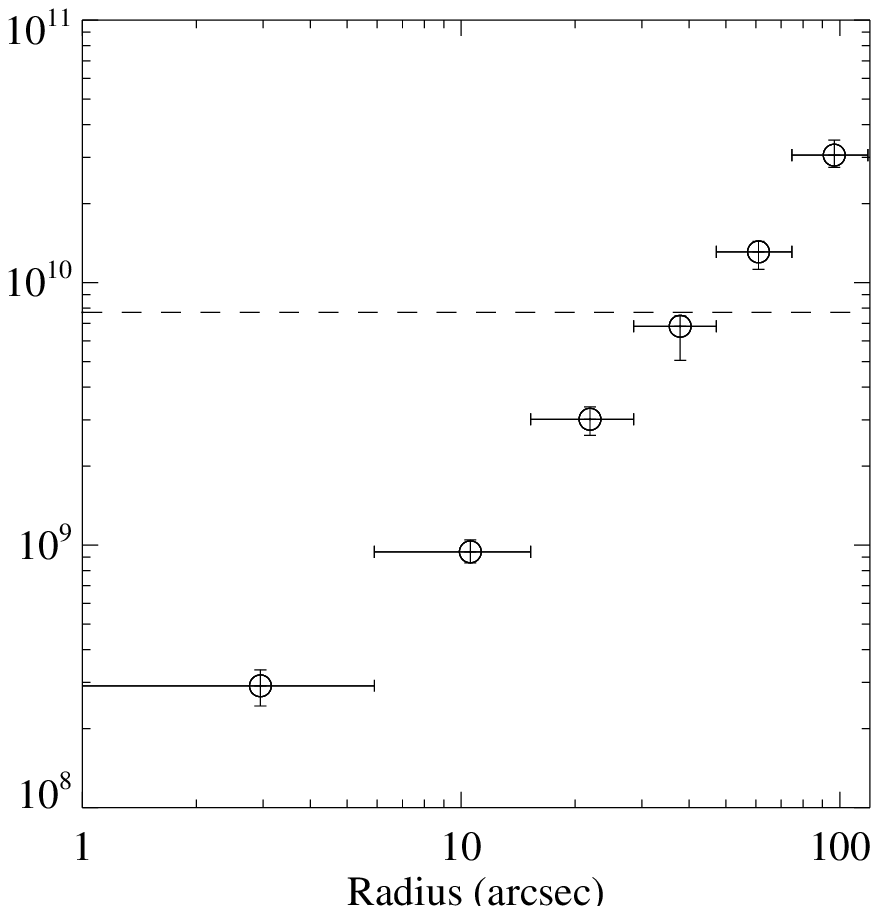}{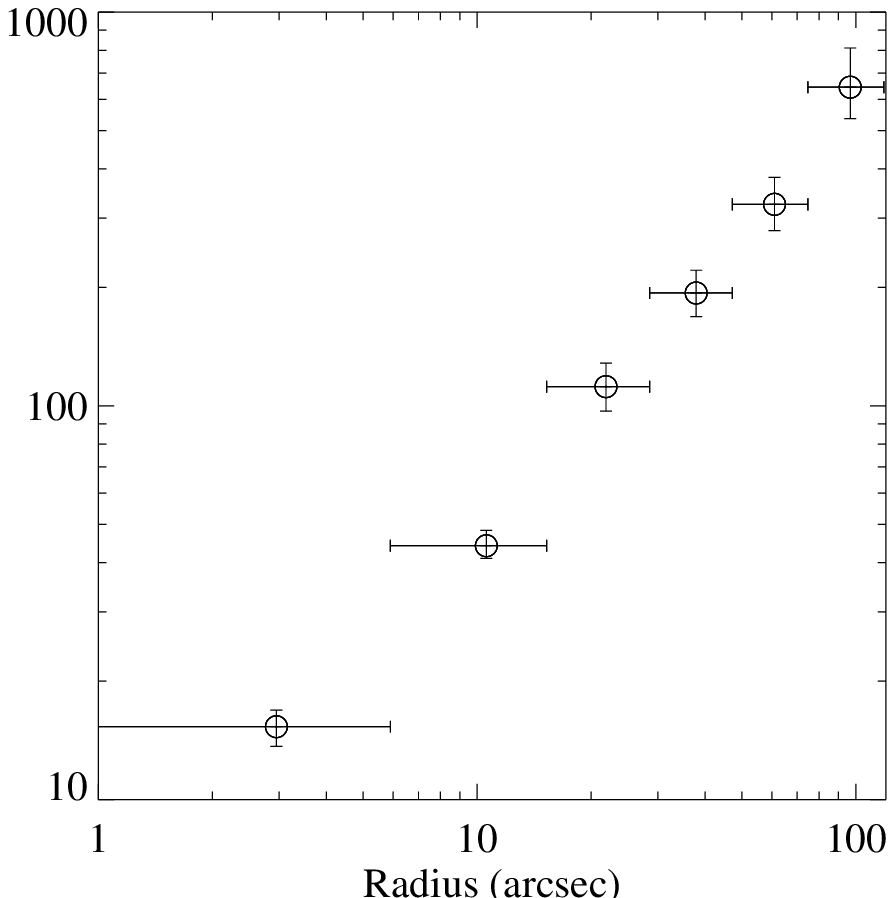}
\caption{\emph{Left:} Cooling time profile for the hot gas. The dashed line represents a cooling time of $7.7 \times 10^9$ yr,
which is the look-back time to a redshift of one, which we assume to be roughly the epoch of cluster formation. This corresponds
to a cooling radius os 41 arcsec or 161 kpc \emph{Right:} Entropy profile of the hot gas.
\label{fig-xray-image}}
\end{figure}

\clearpage

\begin{figure}
\plotone{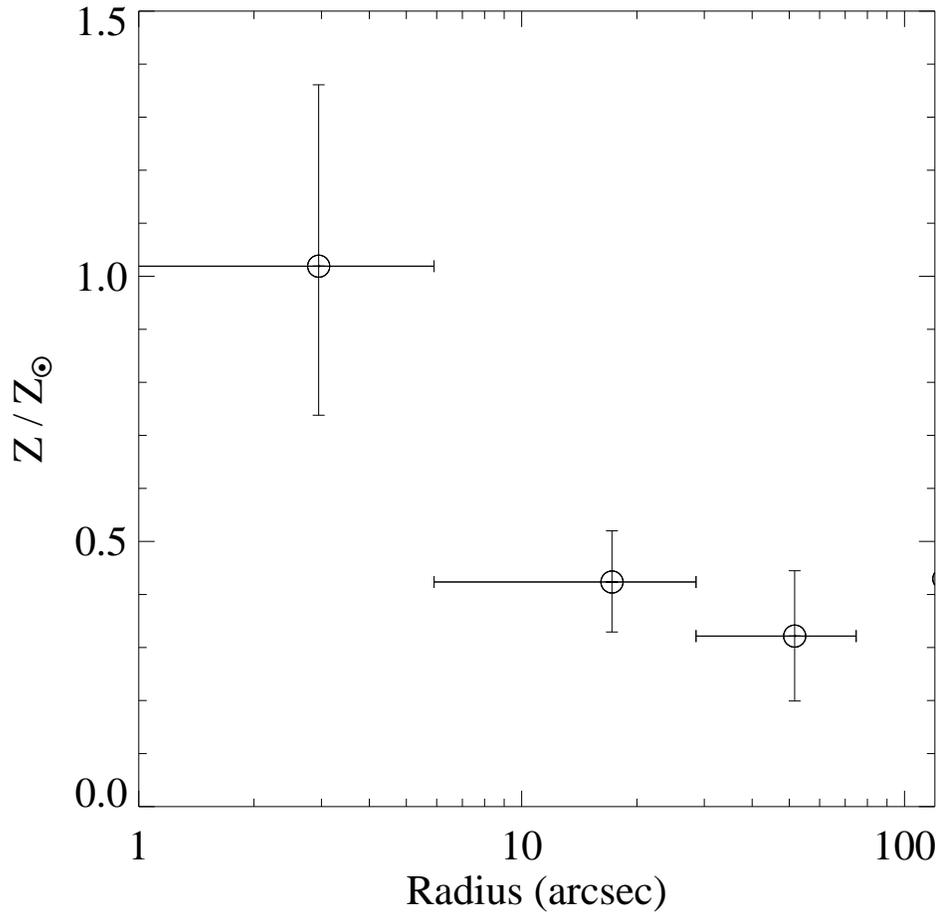}
\caption{Abundance profile of the hot gas in solar units.}
\end{figure}

\clearpage

\begin{figure}[h]
\hbox{
\hspace{0.1in}
\psfig{figure=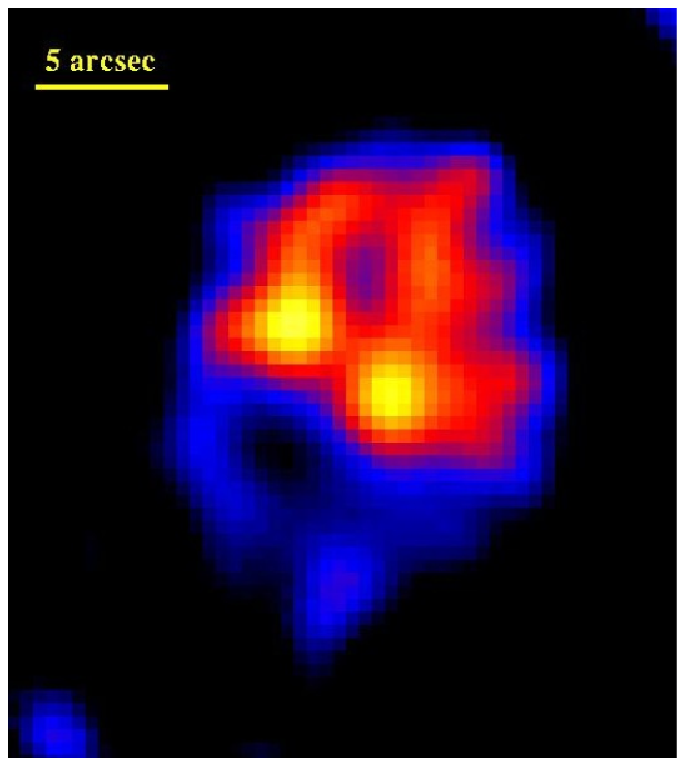,height=3.0in,width=3.in}
\psfig{figure=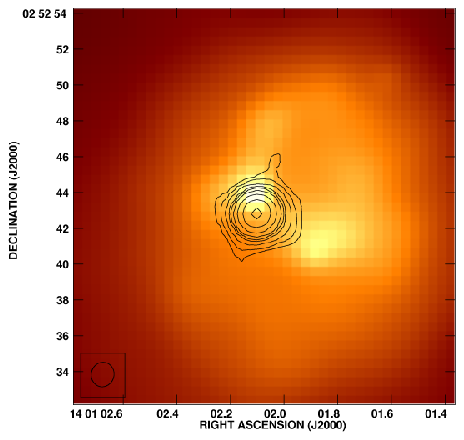,height=3.0in,width=3in}
}
\begin{minipage}[h]{5.0truein}
\vspace{0.1in}
Fig.9 -- \emph{left:} 66.5 ksec image of the core of the cluster taken
recently with the ACIS-I detector (Wise et al. 2006, private communication)
showing the pair of cavities to the north west and
south east of the nucleus, which lies between the bright spots of emission
near the center of the picture. \emph{Right:} Shorter ACIS-S X-ray image 
on which the X-ray analysis was done showing the radio source superposed.
\end{minipage}
\end{figure}

\begin{deluxetable}{lccccccc}
\tabletypesize{\scriptsize}
%\rotate
\tablecaption{Starburst Properties}
\tablewidth{0pt}

\tablehead{

\colhead{  Cluster                                           }& 
\colhead{  SFR                                               }& 
\colhead{  $M_{\rm gas}$                                     }&
\colhead{  $R_{\rm burst}$                                   }& 
\colhead{  ${\rm log}\Sigma_{\rm SFR}$                       }& 
\colhead{  ${\rm log}\Sigma_{\rm gas}$                       }&
\colhead{  $\tau_{\rm dyn}^{\rm a)}$                         }&
\colhead{  ${\rm log}\Sigma_{\rm gas}/\tau _{\rm dyn}$       } \\
\colhead{   \nodata                                          }& 
\colhead{  ${\rm M}_\odot~{\rm yr^{-1}}$                     }& 
\colhead{  ${\rm M}_\odot$                                   }&
\colhead{  kpc                                               }& 
\colhead{  ${\rm M}_\odot~{\rm yr^{-1}~kpc^{-2}}$            }& 
\colhead{  ${\rm M}_\odot~{\rm pc^{-2}}$                     }&
\colhead{  ${\rm yr}$                                        }&
\colhead{  ${\rm M}_\odot~{\rm yr^{-1}}~{\rm pc^{-2}}$       }

}

\startdata
Abell 1835 & $100-180$ & $9\times 10^{10}$ & 30 & $-1.19$ & 1.51 & $6.6 \times 10^8$ & $2.14^{\rm a}$ \\
Abell 1068 & $20-70 $ & $4\times 10^{10}$ & 10 & $-0.65$ & 2.12 & $2.2 \times 10^8$ & $2.25^{\rm a}$ \\
\enddata

\tablenotetext{a}{assumes a stellar velocity dispersion of $281~{\rm km~s^{-1}}$ (B\^{\i}rzan et al. 2004)}

\end{deluxetable}

\begin{deluxetable}{cccccccc}
\tabletypesize{\scriptsize}
%\rotate
\tablecaption{Cavity Properties}
\tablewidth{0pt}

\tablehead{
\colhead{Cavity }&
\colhead{$R$ }&
\colhead{$a$ } &
\colhead{$b$ } &
\colhead{$p$ } &
\colhead{$kT$ } &
\colhead{$pV$ } &
\colhead{$t$ }  \\
\colhead{...}  &
\colhead{kpc} &
\colhead{kpc} &
\colhead{kpc} &
\colhead{$10^{-9}~{\rm erg~cm}^{-3}$ } &
\colhead{keV} &
\colhead{$10^{59}$ erg }&
\colhead{Myr}
}
\startdata
North-west &23.3 & 15.5 & 11.6 & $1.0\pm 0.1$ & $4.3\pm 0.4$& $2.6^{+2.9}_{-1.0}$ & 41\\
South-east &16.6 & 13.6 & 9.7  & $1.1\pm 0.1$ & $3.8\pm 0.3$& $1.7_{-0.3}^{+1.8}$ & 27\\ 
\enddata

\end{deluxetable}

\end{document}